\documentstyle[psfig]{mn}

\title{Principal Component Analysis for Spectral Indices of Stellar 
Populations}

\author[X. Kong \& F.Z. Cheng]{X. Kong$^{1,2}$ and F. Z. 
Cheng$^{1,2}$\\
$^1$ Center for Astrophysics, University of Science and Technology 
of China, 230026, Hefei, P. R. China \\
$^2$ National Astronomical Observatories, Beijing Astrophysics 
Center, Chinese Academy of Sciences, 100012, Beijing, P. R. China\\}

\date{Accepted . Received. }

\begin{document}

\maketitle

\label{firstpage}

\begin{abstract}

We apply the method of principal component analysis to a sample of 
simple stellar population to select some age sensitive spectral 
indices.  Besides the well-known age sensitive index, H$\beta$, we 
find some other spectral indices have great potential to determine 
the age of stellar population, such as G4300, Fe4383, C$_2$4668, 
and Mg$b$.  In addition, we find these spectral indices sensitivity 
to age depends on the metallicity of SSP, H$\beta$ and G4300 are 
more suited to determine the age of low metallicity stellar 
population, C$_2$4668 and Mg$b$ are more suited to the high 
metallicity stellar population.  The results suggest that 
principal component analysis method provides a more objective and 
informative alternative to diagnostics by individual spectral 
lines.

\end{abstract}


\section{Introduction}

To understand how galaxies are formed and evolve we need to 
investigate their stellar populations and to derive their main 
parameters, such as metallicity and age. This study plays an 
important role in our understanding of galaxy properties (Kennicutt 
1998; Maraston 1998; Padoan et al. 1997).  During the last decade 
particularly using population synthesis method has performed 
determination of galaxy properties.  It has been used extensively 
by many authors for all kinds of galaxies (Leitherer et al.\ 1996). 
However, key parameters in interpretation of the observed 
properties are the metallicity and the age. The problem is the 
degeneracy of the effects from variations in age and in metallicity, 
even in the simplest unresolved stellar systems their effects are 
very difficult to separate (Vazdekis et al. 1997). It makes the 
determination of age and metallicity uncertain.  To determine 
the age and metallicity more accurately, strong efforts have been 
devoted to select some spectral features that are more sensitive 
to age, and others are more sensitive to metallicity. In 1994, G. 
Worthey developed a very efficient method to investigate the age 
and metallicity effects in the integrated light of stellar 
populations.  In his method, suppose an index varies by an amount 
$\Delta I$, it can be explain $\Delta I$ by either a pure metallicity 
variation or a pure age variation. The Z sensitivity parameter is 
the ratio of the percentage change in age to the percentage change 
in Z, with larger numbers indicating greater metallicity 
sensitivity (Worthey 1994). Using this method, some metallicity 
sensitive spectral indices are found, but only two age sensitive 
spectral indices are found.

In this paper, we apply a different statistical technique, principal 
component analysis (PCA) to simple stellar population sample, to 
investigate on the reliability of the relations between spectral 
indices and age.  We want to extract other spectral indices that 
can be used to determine the age of stellar population.  The samples 
of simple stellar population (SSP) spectra come from Bressan et al.\ 
(1996) and Bruzual \& Charlot (1996).  For each spectra, 21 spectral 
indices in the Lick/IDS system be measured first (Trager et al.\ 
1998), then we use PCA method to find some spectral indices that 
are more sensitive to the age. We leave the synthetic galaxy spectral 
indices with different star formation history and observational 
galaxy spectral indices to a future paper. However we already notice 
an encouraging good resemblance between the results of PCA and those 
one gets from observation (Worthey 1994).

The organization of the rest of this paper is as follows.  In 
section~2 we present PCA method.  In section~3 we describe the 
definitions of spectral indices, the synthetic models we have used.  
In section~4 we analyze a sample of SSP spectra with different ages, 
find the first principal components (PCs), discuss the significance 
of different spectral indices to determine the age, and how 
different age sampling affects the results.  In section~5, we 
illustrate that the uncertainties will not affect our results, and 
compare our results with previously published values. The 
conclusions are summarized in section~6.

\section{Principal Component Analysis}

Principal component analysis is known to be powerful technique for 
unveiling correlation between variables in a data set and for 
determining the intrinsic dimensionality of a parameter space 
(Connolly \& Szalay 1999).  It is an orthogonal transformation that 
allows the building of more compact linear combinations of data that 
are optimal with respect to the mean square error criterion.  The 
algorithm calculates a new base with the minimum set of orthogonal 
axes that describes the observed variance of the data. Finding the 
new axes is an eigenvalue formation matrix between the old base and 
the new base.  The new orthogonal basis is composed of vectors 
called principal components (PCs).  The corresponding eigenvalues 
represent the variances of the parameters in the new bases (Francis 
\& Wills 1999).

The formulation of standard PCA is as follows (see e.g. Murtagh \& 
Heck 1987). Consider a set of $n$ objects ($i=1,~n$), each with $m$ 
parameters ($j=1,~m$). If $x_{ij}$ are the original measurements, 
the starting point in PCA method is an observation matrix $X = 
(x_{ij})_{n \times m}$ in which row vectors list observations 
characterizing an object.

We then construct a covariance matrix  $C = (c_{ij})_{m \times m}$, 
where

\begin{equation}
c_{jk}={1\over n-1 }\sum_{i=1}^n
(x_{ij}-\overline{x}_j)(x_{ik}-\overline{x}_k)
\quad 1\le j, k \le m
\end{equation}


where $\overline{x}_j$ and $\overline{x}_k$ are the mean of column 
vector.

It can be shown that the axis along which the variance is maximal 
is the eigenvector $\bmath{e_1}$ of the matrix equation

\begin{equation}
C\bmath{e_1}=l_1\bmath{e_1,}
\end{equation}
where $l_1$ is the largest eigenvalue of matrix C, which is in fact 
the variance along the new axis.  The other principal axes and 
eigenvectors obey similar equations. It is convenient to sort them 
in decreasing order, and to quantify the fractional variance by 
$l_j/\sum_j l_j$. The matrix of all the eigenvectors forms a new 
set of orthogonal axes that are ideally suitable to a description 
of the data set.  If the main variance of the data set lies in a 
small dimensional space, then one can get a good visualization of 
it by plotting the data projected on the first few eigenvectors (PCs). 
The projection of a vector $\bmath{x}$ on the eigenvector 
$\bmath{e_j}$ being $\bmath{x \cdot e_j}$, where $\bmath{x}$ is a 
row vector of $X$.  The fractional variance of the first PCs tells 
us to what extent the data lie in a given low dimensional space (Ronen 
et al. 1999).

In what follows, we will apply this technique to different spectral 
index samples. To avoid confusion we denote eigenvectors by PC1, 
PC2 etc., and projections on the new axes by pc1, pc2 etc. For the 
$k$th object, $pc_j = \bmath{x}\cdot \bmath{e_j}=e_{j1} x_{k1} + 
\cdots + e_{ji} x_{ki} + \cdots + e_{jm} x_{km}$.

\section{Indices of Simple Stellar Population}

Among the libraries of spectral indices, the most extensive one is 
created by Faber and coworkers at Lick Observatory, the Lick/IDS 
System (Faber et al.\ 1985; Worthey et al.\ 1994; Trager et al.\ 
1998). In the Lick/IDS system, absorption-line strengths are 
described by ``indices,'' in which a central ``feature'' band-pass 
is flanked to the blue and red by ``pseudo-continuum'' band-passes.  
It is fully described in Worthey et al.\ (1994) and Trager et al.\ 
(1998).  Table 2\footnote{This table is available electronically 
from the Astrophysical Data Center 
(http://adc.gsfc.nasa.gov/adc.html).} of Trager et al. (1998) 
presents the band-passes of these 21 Lick/IDS absorption line 
indices and the features measured by these indices.

Simple stellar population is defined as a single generation of 
coeval stars with fixed parameters such as metallicity, initial mass 
function, etc (Buzzoni 1997).  To find some spectral indices that 
can be used to determine the age of stellar population, as the first 
step, we apply PCA method to SSP only.  To assess how model-
dependent our conclusions are, we use the spectral indices of SSP 
form two different synthetic models.

The first SSP sample is the one used by Tantalo et al. (1996).  
Detailed descriptions of the SSP model can be found in Bressan et 
al.  (1994), Silva (1995), and Tantalo et al. (1996).  These SSPs 
extend the ZAMS to 0.15 $M_{\odot}$. The rate of mass loss during 
the RGB and TP-AGB phases is according to the Reimers (1975) 
relationship with $\eta=0.45$. They are computed with the new 
library of stellar spectra described in Silva (1995), and Tantalo 
et al. (1996). These SSPs are shortly referred to as SIL-SSP. The 
spectral indices of this sample are calculated by means of the 
empirical calibrations of Worthey (1992) and Worthey et al. (1994).

The second one is isochronal synthesis model of Bruzual \& Charlot 
(1996). It has extended the Bruzual \& Charlot (1993) evolutionary 
population synthesis model. The updated version provides the 
evolution of the spectrophotometric properties for a wide range of 
stellar metallicity.  The stellar evolution tracks are from Padova 
group (Bressan et al. 1993).  The stellar spectra library is from 
Lejeune et al. (1997,1998).  These SSPs are shortly referred to as 
BC-SSP.

The spectral indices in Lick/IDS system for SIL-SSP and BC-SSP are 
published in the AAS CD-ROM Series, Vol. 7.  The technique to 
calculate spectral indices of SSP is amply described in Worthey 
(1994), Bressan et al.(1996), Tantalo et al. (1998) and Longhetti 
et al. (1998).

\section{PCA to SSP}

To find some spectral indices that are more sensitive to the age, 
we will apply PCA method to SSPs in this section.  

\subsection{PCA to SIL-SSP}

Given an IMF and metallicity, evolutionary synthesis model provides 
SSP's spectra as a function of age in the sample of SIL-SSP.  Each 
data set contains about 50 SSPs, corresponding to 50 time steps from 
0.004 to 20 Gyr. Each SSP has 1206 wavelength points in the range 
from 91 {\AA} to 100 $\mu m$ (Leitherer et al. 1996).  Because the 
spectral indices of CN$_1$, CN$_2$ and TiO$_2$ are equal to 0.0 or 
very little for young stellar population, and there are some 
uncertainties at high $Z$ SSP, so we perform PCA on a sample of SSP with 
age 1--15 Gyr, and Z=0.0004 to 0.05.

For our first study, we performed PCA on the SIL-SSP with 
$Z=Z_\odot=0.02$.  The results are shown in Table 1.  Column 
(2)--(3) show the first 2 principal components (PCs) out of a total 
of 21 principal components.  The 3{\it rd} row in Table 1 gives the 
variances (eigenvalues) of the data along the direction of the 
corresponding principal component.  By convention, the principal 
components are given in order of their contribution to the total 
variance, it is given as `Proportion' in the 4{\it th} row.  The 
columns of numbers for each principal component represent the 
weights assigned to each input variable.  Thus pc1 $= 0.02\times 
x_1 + 0.01\times x_2 + 0.14 \times x_3 + 0.49 \times x_4 + 0.49 \times 
x_5 $ \ldots, where $x_1$, $x_2$, $x_3$, $x_4$, and $x_5$ \ldots, 
are the values of the normalized variables corresponding to CN$_1$, 
CN$_2$, Ca4227 ,G4300 and Fe4383, etc. If the value in Column 2 of 
Table 1 is large, it suggests that the corresponding spectral index 
is important to PC1. The first principal component is elongated with 
variance 5.11, and accounts for about most the total variance. It 
is therefore likely to be highly significant. In SIL-SSP sample, 
only age is variable, so PC1 shows the age information, it correlates 
with ages.  We can use PC1 to determine the age of a SSP.

\begin{table*}
\caption{Results of Eigenanalysis for SIL-SSP (Padova group): 
Percentage of variance explained by the principal components and 
Weights on the first two components for each index involved in the 
analysis. See Trager (1998) for index definition.\label{tab1}}

\begin{center}
\begin{tabular}{ll|rrrrrrrrrr}
\hline
\multicolumn{2}{l}{Metallicity}
&\multicolumn{2}{|c}{Z=0.02}
&\multicolumn{2}{c}{Z=0.0004}
&\multicolumn{2}{c}{Z=0.004}
&\multicolumn{2}{c}{Z=0.008}
&\multicolumn{2}{c}{Z=0.05}\\
\hline
\multicolumn{2}{l}{Index}&PC1&PC2&PC1&PC2&PC1&PC2&PC1&PC2&PC1&P
C2 \\
\multicolumn{2}{l}{Eigenvalue}&5.11& 0.02 & 2.86& 0.02 & 5.25&
0.03 & 5.09& 0.02 & 5.06& 0.04\\
\multicolumn{2}{l}{Proportion}&99.42\% & 0.47\% & 99.01\% & 0.63\%
& 99.35\% & 0.51\% & 99.39\% & 0.40\% & 99.04\% & 0.78\% \\
\\
\multicolumn{2}{l}{Variable}&PC1&PC2&PC1&PC2&PC1&PC2&PC1&PC2&PC
1&PC2
\\
\hline
01& CN$_1$   &0.02& 0.02 & 0.02&-0.02 & 0.01& 0.01 & 0.01& 0.01 &
0.02& 0.04\\
02& CN$_2$   &0.01& 0.02 & 0.01&-0.01 & 0.01& 0.01 & 0.01& 0.02 &
0.02& 0.05\\
03& Ca4227   &0.14& 0.25 & 0.06& 0.13 & 0.08&-0.21 & 0.10&-0.22 &
0.18& 0.26\\
04& G4300    &0.49&-0.53 & 0.79& 0.05 & 0.67& 0.18 & 0.60&-0.01 &
0.35&-0.70\\
05& Fe4383   &0.49& 0.39 & 0.33& 0.16 & 0.44&-0.29 & 0.46&-0.34 &
0.50& 0.22\\
06& Ca4455   &0.12& 0.02 & 0.04& 0.04 & 0.10&-0.03 & 0.11& 0.00 &
0.13& 0.01\\
07& Fe4531   &0.17& 0.05 & 0.22&-0.16 & 0.19& 0.12 & 0.18& 0.13 &
0.17& 0.07\\
08& C$_2$4668&0.36&-0.10 &-0.06& 0.37 & 0.22& 0.04 & 0.29& 0.36 &
0.42&-0.11\\
09& H$\beta$&-0.32& 0.33 &-0.41& 0.22 &-0.39&-0.08 &-0.37&-0.16
&-0.27& 0.25\\
10& Fe5015   &0.19&-0.27 & 0.16&-0.16 & 0.19& 0.50 & 0.20& 0.55 &
0.19&-0.25\\
11& Mg1      &0.01& 0.02 & 0.00& 0.02 & 0.01&-0.02 & 0.01&-0.01 &
0.01& 0.02\\
12& Mg2      &0.02& 0.02 & 0.00& 0.03 & 0.01&-0.03 & 0.01&-0.02 &
0.02& 0.02\\
13& Mg$b$    &0.28& 0.31 & 0.07& 0.65 & 0.16&-0.33 & 0.22&-0.29 &
0.34& 0.22\\
14&Fe5270    &0.15& 0.07 & 0.08& 0.03 & 0.13& 0.01 & 0.14& 0.09 &
0.15& 0.07\\
15&Fe5335    &0.14& 0.06 & 0.09& 0.08 & 0.12&-0.15 & 0.14& 0.06 &
0.15& 0.09\\
16&Fe5406    &0.11& 0.06 & 0.03& 0.09 & 0.08& 0.00 & 0.09& 0.08 &
0.12& 0.03\\
17&Fe5709    &0.04&-0.01 & 0.01&-0.01 & 0.04& 0.06 & 0.04& 0.09 &
0.03&-0.01\\
18&Fe5782    &0.03&-0.01 & 0.00& 0.03 & 0.03& 0.00 & 0.03& 0.10 &
0.04& 0.01\\
19&NaD     &0.22& 0.45 & 0.03& 0.53 & 0.11&-0.65 & 0.15&-0.49 &
0.29& 0.43\\
20&TiO$_1$ &0.00& 0.00 & 0.00& 0.01 & 0.00&-0.01 & 0.00& 0.00 &
0.00& 0.00\\
21&TiO$_2$ &0.00&-0.01 & 0.00& 0.02 & 0.00&-0.01 & 0.00& 0.01 &
0.01&-0.02\\
\hline
\end{tabular}
\end{center}
\end{table*}

This result shows that most of these spectral indices correlate with 
PC1, but the correlation involving G4300, Fe4383, C$_2$4668, 
H$\beta$, and Mg$b$ are very strong.  So these spectral indices can 
be used to determine the age of stellar system.  PC2 accounts for 
only 1\% of the variance in this data. Some spectral indices appear 
to contribute to both PC1 and PC2, but the contribution of PC2 is 
very little, therefore any spectral indices present in PC2 represent 
only fine tuning to the main variation by which PC1 represents. PC2 
is not a significant component.

To inspect how sensitive the results of PCA method are to the 
different metallicity, we performed PCA on the SIL-SSP with other 
metallicity. The results of performing PCA on these SSPs are shown 
in the last 8 columns of Table 1. There are some distinct characters 
in the results.  First, we find that the spectral indices G4300, 
Fe4383, C$_2$4668, H$\beta$, and Mg$b$ are very strong correlated 
with PC1 also. Second, the contribution from PC1 is much more 
important than that from PC2. Third, we find that the spectral 
indices contributing to the PCs are different from the SSPs with 
different metallicity. Mg$b$ and C$_2$4668 contributing to low 
metallicity SSP are small, they cannot fit to determine the age for 
low metallicity SSP but for high metallicity SSP. On the contrary, 
G4300 and H$\beta$ are more fit to determine the age for low 
metallicity SSP.  Actually, the sodium index Na$D$ is also an 
important index, with respect to the PC1. We did not select it due 
to the considerable effect of the interstellar sodium line as 
previously reported(Worthey 1994)

\subsection{PCA to BC-SSP}

There exist a number of SSP models that are synthesized with 
different approaches.  It is important to check the sensitivity of 
our results to the SSP models adopted.  The Bruzual \& Charlot's 
model is suitable for the comparison, because it uses a similar 
technique of ``isochronal synthesis'' to predict the spectral 
evolution of stellar population. Each data set (metallicity fix) 
contains about 221 SSPs, corresponding to 221 time steps from 0 to 
20 Gyr. Each SSP has 1206 wavelength points in the range from 5 {\AA} 
to 100 $\mu m$ (Bruzual \& Charlot 1996).  Similar to SIL-SSP, we 
performed PCA method to the stellar population (1--15 Gyr) with 
metallicity Z=0.0004 to 0.05. The results are shown in Table 2.

\begin{table*}
\caption{Results of Eigenanalysis for BC-SSP (Bruzual \& Charlot
1996)
\label{tab2}}
\begin{center}
\begin{tabular}{ll|rrrrrrrrrr}
\hline
\multicolumn{2}{l}{Metallicity}
&\multicolumn{2}{c}{Z=0.0004}
&\multicolumn{2}{c}{Z=0.004}
&\multicolumn{2}{c}{Z=0.008}
&\multicolumn{2}{c}{Z=0.02}
&\multicolumn{2}{c}{Z=0.05}\\
\hline
\multicolumn{2}{l}{Index}&PC1&PC2&PC1&PC2&PC1&PC2&PC1&PC2&PC1&P
C2 \\
\multicolumn{2}{l}{Eigenvalue}&1.17&0.02& 2.68& 0.04 & 3.39 &0.02 
&
4.66& 0.02 & 5.06& 0.06 \\
\multicolumn{2}{l}{Proportion}&98.25\%&1.37\%&98.30\%&1.56\%&
99.16\% &0.68\%&
99.54\%&0.37\%&98.73\%&1.24\%\\
\\
\multicolumn{2}{l}{Variable}&PC1&PC2&PC1&PC2&PC1&PC2&PC1&PC2&PC
1&PC2
\\
\hline
01&CN$_1$    & 0.00& 0.00 & 0.00& 0.00 & 0.00&-0.01 & 0.01&-0.02 
&
0.01& 0.05\\
02&CN$_2$    & 0.00& 0.00 & 0.00&-0.01 & 0.00&-0.01 & 0.01&-0.03 
&
0.02& 0.06\\
03&Ca4227    & 0.10& 0.17 & 0.12&-0.03 & 0.12& 0.02 & 0.13&-0.28 
&
0.17& 0.20\\
04&G4300     & 0.34& 0.17 & 0.57& 0.33 & 0.54& 0.34 & 0.48& 0.68
& 0.32&-0.81\\
05& Fe4383   & 0.18& 0.15 & 0.39& 0.37 & 0.46& 0.40 & 0.51&-0.18
& 0.47&-0.03\\
06&Ca4455    &-0.07& 0.12 & 0.13&-0.04 & 0.12&-0.09 & 0.11&-0.09 
&
0.11& 0.01\\
07&Fe4531    &-0.23& 0.26 & 0.22&-0.23 & 0.19&-0.25 & 0.18&-0.12 
&
0.18& 0.07\\
08& C$_2$4668&-0.01& 0.27 & 0.18& 0.12 & 0.29&-0.01 & 0.38&-0.07 
&
0.51& 0.24\\
09& H$\beta$ &-0.83& 0.45 &-0.42& 0.70 &-0.32& 0.64 &-0.26&-0.16
&-0.22& 0.14\\
10&Fe5015    &-0.21& 0.41 & 0.24&-0.29 & 0.24&-0.36 & 0.21& 0.10
& 0.21&-0.13\\
11&Mg1       & 0.00& 0.01 & 0.01& 0.00 & 0.01& 0.00 & 0.01&-0.03 
&
0.01& 0.01\\
12&Mg2       & 0.01& 0.03 & 0.02& 0.01 & 0.02& 0.00 & 0.02&-0.04 
&
0.02& 0.02\\
13&Mg$b$     & 0.11& 0.44 & 0.28& 0.28 & 0.31& 0.18 & 0.31&-0.11 
&
0.32& 0.04\\
14&Fe5270    & 0.11& 0.12 & 0.17&-0.10 & 0.16&-0.17 & 0.15&-0.15 
&
0.15& 0.10\\
15&Fe5335    & 0.11& 0.24 & 0.16&-0.03 & 0.16&-0.17 & 0.15&-0.14 
&
0.16& 0.06\\
16&Fe5406    & 0.05& 0.07 & 0.11&-0.02 & 0.10&-0.11 & 0.11&-0.14 
&
0.12& 0.04\\
17&Fe5709    & 0.03&-0.01 & 0.03&-0.07 & 0.03&-0.05 & 0.04&-0.03
& 0.03&-0.01\\
18&Fe5782    & 0.02&-0.01 & 0.04&-0.05 & 0.03&-0.09 & 0.03&-0.05 
&
0.04& 0.04\\
19&NaD       & 0.10& 0.35 & 0.16& 0.13 & 0.18&-0.03 & 0.20&-0.52 
&
0.28& 0.42\\
20&TiO$_1$   & 0.00& 0.01 & 0.00& 0.00 & 0.00&-0.01 & 0.00& 0.00 
&
0.00& 0.00\\
21&TiO$_2$   & 0.00& 0.02 & 0.01& 0.00 & 0.01&-0.02 & 0.00& 0.00 
&
0.01& 0.00\\
\hline
\end{tabular}
\end{center}
\end{table*}

In Table 2, the first principal component also accounts for about 
most the total variance, more than 98\%. PC2 can be ignored, it maybe 
caused some uncertainties.  This result is consistent with the 
principle of PCA method: One of important application of PCA is known 
to be powerful technique for determining the intrinsic 
dimensionality of a parameter space. Given an IMF and metallicity, 
SSP has one parameter, age.  This parameter can be expressed by PC1. 
From Table 1 and Table 2, we find the results from different SSP 
are very similar, it can be explained that our results are not 
sensitive to the SSP models adopted. The spectral indices, G4300, 
Fe4383, C$_2$4668, H$\beta$, and Mg$b$, can be used to determine 
the age of stellar population.

\subsection{PCA to the SSPs with different metallicity}

To extract an objective information on the age-metallicity 
sensitivity of the narrow band indices, we apply PCA method to the 
whole set of SSPs.  As a result, we find a little information for 
metallicity. We think it can be understood easily, because there 
are only five kind metallicities for the sample of SSP, the 
metallicity point is less than the kind of spectral indices. So we 
can not obtain more metallicity information from this statistical 
technique. In the future paper, we will observe some spectra of 
globular cluster with different metallicity, and perform a 
principal component analysis method to them, It maybe help us to 
get the metallicity sensitive indices.

To overcome this difficulty, we 
have selected 3 spectral indices from Worthey (1994).  There are 
H$\beta$ which is age sensitive, Fe5015 which is metallicity 
sensitive and Ca4227 which isn't age or metallicity sensitive. 
We apply PCA method to these 3 indices, with same age and different 
metallicities(5  kind metallicities).  
The results are shown in Column (3)--(8) of Table 3 for SIL-SSPs, and in Column 
(9)--(14) for BC-SSPs.  The age of SSP is $T=10^9$ yr, $10^{10}$ yr, and 
$1.5\times 10^9$ yr.  The 1{\it st} row in Table 3 gives the age 
of the SSP, the meaning of next 3 rows same as Table 1. The numbers 
in the last 3 rows represent the weights assigned to each input 
variable. If the value in these columns is large, it suggests that 
the corresponding spectral index is important. From this table, we 
find the first principal component accounts for about most the total 
variance. This is therefore to be highly significant.  In this SSP 
sample, since the age of SSP is same, only metallicity is variable, 
so PC1 must show the metallicity information. In the table, the value 
of the normalized variable corresponding to Fe5015 is large, it 
suggests that the corresponding spectral index, Fe5015, is 
important to metallicity. The value of H$\beta$ is small, it seems that 
H$\beta$ is not important to metallicity.  This result is the same 
as the result of Worthey (1994).  From this analysis, we know PCA 
is a very efficient method, it can be used to select some spectral 
indices that will help us to determine the age or metallicity of 
SSP, if the age and metallicity points are enough.

\begin{table*}
\caption{Results of Eigenanalysis for the SSP with different 
metallicity.
\label{tab3}}
\begin{center}
\begin{tabular}{ll|rrrrrr|rrrrrr}
\hline
\multicolumn{2}{l}{}
&\multicolumn{6}{|c|}{SIL-SSP}
&\multicolumn{6}{|c|}{BC-SSP}\\
\cline{4-7}\cline{10-13}\\
\multicolumn{2}{l}{Age}
&\multicolumn{2}{c}{$10^9$ yr}
&\multicolumn{2}{c}{$10^{10}$ yr}
&\multicolumn{2}{c|}{$1.5\times 10^9$ yr}
&\multicolumn{2}{c}{$10^9$ yr}
&\multicolumn{2}{c}{$10^{10}$ yr}
&\multicolumn{2}{c}{$1.5\times 10^9$ yr}
\\
\hline
\multicolumn{2}{l}{Index}&PC1&PC2&PC1&PC2&PC1&PC2&PC1&PC2&PC1&P
C2&PC1&PC2\\
\multicolumn{2}{l}{Eigenvalue}&2.96  &0.03   
&3.87&0.02&4.53&0.06&
3.35&0.02&4.44&0.01&4.69&0.01\\
\multicolumn{2}{l}{Proportion}&98.8\%&1.16\% 
&99.5&0.52&98.8&1.20&
99.3&0.66&99.8&0.13&99.7&0.27\\
\\
\multicolumn{2}{l}{Variable}&PC1&PC2&PC1&PC2&PC1&PC2&PC1&PC2&PC
1&PC2&PC1&PC2
\\
\hline

03&Ca4227    & 0.17 &-0.15 &0.34&0.91&0.32&0.56&
0.16&0.21&0.28&0.94&0.31&0.95\\
09& H$\beta$ &-0.32 & 0.93 &-0.21&0.32&-0.30&0.83&
-0.50&0.86&-0.19&0.24&-0.17&0.05\\
10&Fe5015    & 0.93 & 0.35 &0.92&-0.26&0.90&0.07&
0.85&0.46&0.94&-0.23 &0.93&-0.31\\
\hline
\end{tabular}
\end{center}
\end{table*}

\subsection{PC1 and Index with Age}

For each SSP sample (metallicity fix), we can obtain $\vec{e_j}$ 
from PCA method. Using $pc_j = \vec{x}\cdot \vec{e_j}$, we can 
calculate the projection of a vector $\vec{x}$ on the eigenvector.  
We plot the projection on the PC1 (pc1) versus age for the 
metallicity range from Z=0.0004 to Z=0.05 in Fig. 1.  Each line 
pattern represents a different metallicity, as indicated inside the 
frame. In Fig. 1(a), we plot the pc1 versus age for SIL-SSP. In Fig. 
1(b), we plot for BC-SSP.  In these figures, the following remarks 
can be made.  (1) It is apparent that there is a uniform tendency 
for the values of pc1 to become larger as the metallicity increases 
from $Z=0.0004$ to $Z=0.05$. (2) These mainly are monotonically 
increasing curve with age, clearly. Therefore, once we know the 
metallicity, we can use this pc1 to determine the age of stellar 
population.  (3) Some small non-monotonic irregularities in the 
curve are probably due to abrupt transition of stars to different 
evolutionary stages. They may be real, but they may also occur if 
the model does not contain enough evolutionary tracks in a certain 
range. In any case, we shall see that these irregularities will be 
smoothed out when considering longer and more realistic star 
formation histories instead of the instantaneous bursts considered 
here.

\begin{figure}
\psfig{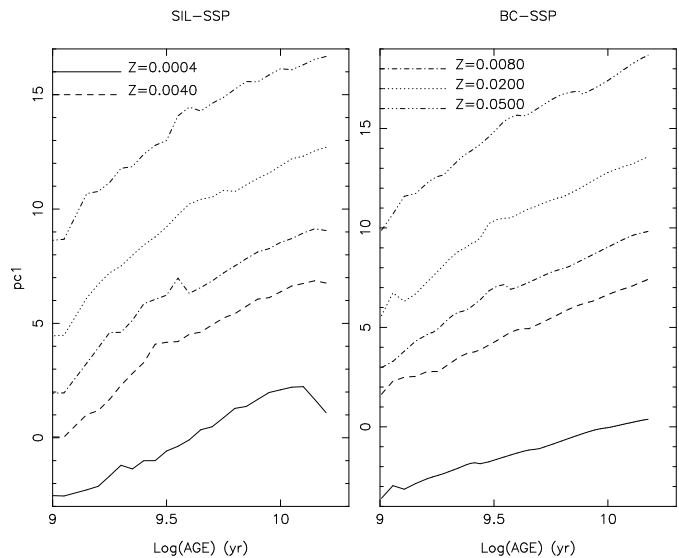}
\caption{
Projection of the spectral indices on the PC1 (pc1) versus
age. The synthetic spectral indices of SSP formed in a single 
instantaneous burst. a) SIL-SSP, b) BC-SSP.}
\label{fig1}
\end{figure}

As an example, we plot the spectral indices with age for SIL-SSP 
with Z=0.02 in figure~2.  We find that spectral indices become 
stronger with increasing age, while Balmer line, H$\beta$, become 
weaker. The change of G4300, Fe4383, C$_2$4668, H$\beta$, and Mg$b$ 
with age are very strong, especially for G4300 and H$\beta$.  These 
results, from Fig. 1 and Fig. 2, consistent with the results from 
PCA method.  So these spectral indices can be used to determine the 
age of stellar system.  The results suggested that PCA is 
efficacious method for find the age sensitive indices.

\begin{figure}
\psfig{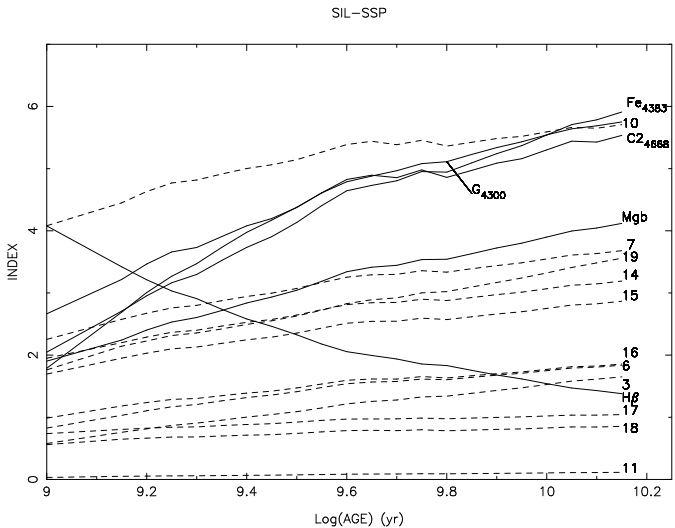}
\caption{ Spectral indices of the SIL-SSP versus age. The solid 
lines show the spectral indices that can be used to determine the 
age of stellar population. We use the spectral index name or serial 
number (in Tab. 1) to indicate it. Some index is not plotted in this 
figure for its change little.}
\label{fig2}
\end{figure}

\section{Discussion}

In this paper, as the first step, we only performed PCA to SSP sample. 
There are two reasons.  Firstly, SSP is simple and reasonably well 
understood, so it is important to see what one can learn by using 
this simplest assumption, and then check whether more complex star 
formation histories give qualitatively similar conclusions. This 
is a common approach often taken in the evolutionary population 
synthesis models for galaxies (Vazdekis et al. 1997).  Secondly, 
the selected spectral indices in this paper can be used to determine 
the age of star cluster. 

The synthesis spectral indices are computed from the SSP of various 
ages and metallicity, under some fitting functions. The 
uncertainties in the fitting function, the stellar evolution input 
and the evolutionary population synthesis computational procedure 
may be affecting our result. But for the result of Maraston et al. 
(1999), the discrepancies introduced by different fitting function 
are relatively small. Form the result state above , the difference 
between the SIL-SSP and BC-SSP is very small, so the uncertainties 
due to the use of different evolutionary population synthesis 
procedures and stellar evolution input can be ignored.  

Compared with the previous methods, PCA method provides a more 
objective and informative alternative to diagnostics by individual 
spectral lines, the PCs represent all the spectral features {\it 
simultaneously}.  Same as Worthey (1994), we also find G4300 and 
H$\beta$ are age sensitive indices that are useful in determining 
age. In addition, we also find some other indices that can be used 
to determine the age of SSP with different metallicities.

\section{Conclusions}

In order to select spectral indices that more sensitive to age than 
metallicity, PCA method has been applied to the Lick indices in the 
population model of SIL-SSP and BC-SSP. It is the first time this 
method is used for selecting the age sensitive spectral indices. 
Below, we summarize our main conclusions.

\begin{enumerate}
\item Using PCA method, we find some spectral indices, such as G4300, 
H$\beta$, C$_2$4668 and Mg$b$, which can be used to determine the 
age of stellar population (1--15 Gyr). These spectral indices will 
help us to determine the age of stellar population more accurate.

\item Important differences are found between the age sensitive 
indices at different metallicities. For example, in low metallicity 
C$_2$4668 and Mg$b$ are not fit to determine the age of stellar 
population, but can be used to determine the age for high metallicity 
stellar population.

\item To understand how sensitive our method is to different stellar 
population synthesis models and metallicities, we have compared the 
results obtained with SIL-SSP and BC-SSP with 5 kind of 
metallicities. We find that although the precise values of PCs are 
difference, the general trend is very similar. We can conclude that 
the uncertainties in the evolutionary population synthesis models 
and metallicities can be ignored.

\end{enumerate}

\section*{Acknowledgments}

We deeply thank the referee, Dr. A. Bressan, for so carefully reading 
an early draft of this paper and for making some very useful comments 
and constructive suggestions, which improved the manuscript.  We 
are indebted to Dr. S. Ronen and Dr. G. Worthey to have made available 
their program. We are grateful to Dr. A. Bressan for providing us 
with a grid of spectral indices for simple stellar population.  This 
work is supported by the Chinese National Natural Science Foundation 
(CNNSF) and Chinese National Pandeng Project.

\end{document}